\documentclass[aps,prb,twocolumn,superscriptaddress,showpacs,floatfix]{revtex4}
\usepackage{graphicx,epsfig}
\usepackage{amssymb,times}
\usepackage{amsmath,amsfonts,bm}
\usepackage{dcolumn}
\graphicspath{{plots/}}
\usepackage{epstopdf}

\def\be{\begin{equation}}
\def\ee{\end{equation}}
\def\etal{{\it et al.}}

\def\g{\gamma}

\newcommand{\av}[1]{\langle #1 \rangle}

\newcommand{\dd}{\mathrm{d}}
\newcommand{\CL}{\mathrm{cl}}

\begin{document}


\title{Dynamical Stability of a Many-body Kapitza Pendulum}
\author{Roberta Citro}
\affiliation{Dipartimento di Fisica ``E. R. Caianiello'' and Spin-CNR, Universita' degli Studi di Salerno,
Via Giovanni Paolo II, I-84084 Fisciano, Italy}
\author{Emanuele G. Dalla Torre}
\affiliation{Department of Physics, Harvard University, Cambridge, MA 02138, USA}
\affiliation{Department of Physics, Bar Ilan University, Ramat Gan 5290002, Israel}
\author{Luca D'Alessio}
\affiliation{Department of Physics, The Pennsylvania State University, University Park, PA 16802, USA}
\affiliation{Department of Physics, Boston University, Boston, MA 02215, USA}
\author{Anatoli Polkovnikov}
\affiliation{Department of Physics, Boston University, Boston, MA 02215, USA}\author{Mehrtash Babadi}
\affiliation{Department of Physics, Harvard University, Cambridge, MA 02138, USA}
\affiliation{Institute for Quantum Information and Matter, California Institute of Technology, Pasadena, CA 91125, USA}
\author{Takashi Oka}
\affiliation{Department of Applied Physics, University of Tokyo,
  Tokyo, 113-8656 Japan }
\author{Eugene Demler}
\affiliation{Department of Physics, Harvard University, Cambridge, MA 02138, USA}
\begin{abstract}
We consider a many-body generalization of the Kapitza pendulum: the periodically-driven sine-Gordon model. We show that this interacting system is dynamically stable to periodic drives with finite frequency and amplitude. This finding is in contrast to the common belief that periodically-driven unbounded interacting systems should always tend to an absorbing infinite-temperature state. The transition to an unstable absorbing state is described by a change in the sign of the kinetic term in the effective Floquet Hamiltonian and controlled by the short-wavelength degrees of freedom. We investigate the stability phase diagram through an analytic high-frequency expansion, a self-consistent variational approach, and a numeric semiclassical calculations. Classical and quantum experiments are proposed to verify the validity of our results.
\end{abstract}

\date{\today}
\maketitle

\def \mysection#1{\section{#1}}


\mysection{Introduction}
Motivated by advances in ultra-cold atoms\cite{eckardt05,lignier07,struck_2011,struck_2012,parker_2013}, the stability of periodically-driven many-body systems is the subject of several recent studies \cite{russomanno_2013A,dalessio_2013,russomanno_2013B,lazarides_2014A,mueller_2014,abanin_2014A,lazarides_2014B,ponte_bis_2014,abanin_2014B}. According to the second law of thermodynamics, isolated equilibrium systems can only increase their energy when undergoing a cyclic process. For many-body interacting ergodic systems, it is often assumed that they will heat monotonously, asymptotically approaching an infinite-temperature state \cite{russomanno_2013B,dalessio_2014,lazarides_2014A,abanin_2014A}. In contrast, for small systems such as a single two-level system (spin), thermalization is not expected to occur and periodic alternations
of heating and cooling (Rabi oscillations) are predicted.
A harmonic oscillator can display a transition between these two behaviors, known as ``parametric resonance'' \cite{LL}: depending on the amplitude and frequency of the periodic drive, the oscillation amplitude either increases indefinitely, or displays periodic oscillations. An interesting question regards how much of this rich dynamics remains when many-degrees of freedom are considered.
%

This question was addressed for example by Russomanno \etal \cite{russomanno_2013A}, who studied the time evolution of the  transverse-field Ising (TI) model. This model is integrable and never flows to an infinite-temperature state. In the translational-invariant case, this result can be rationalized by noting that the TI model is integrable and can be  mapped to an ensemble of decoupled two-level systems (spin-waves with a well defined wavevector), each of whom periodically oscillates in time and never equilibrates. In this sense, the findings of Ref.s~[\onlinecite{abanin_2014A,lazarides_2014B,ponte_bis_2014,abanin_2014B}] on periodically-driven disordered systems  subject to a local driving falls into the same category: many-body localized (MBL) systems are effectively integrable because they can be described as decoupled local degrees of freedom as well\cite{vosk_2014,huse_2013}. When the driving frequency is higher than a given threshold, MBL systems remain localized and no not thermalize. This is in contrast to ergodic systems, which are expected to thermalize to an infinite temperature for any periodic drive\cite{russomanno_2013A,dalessio_2014,lazarides_2014A,abanin_2014A}. These findings are in apparent contradiction to earlier numerical studies \cite{prosen_98,prosen1_98,prosen_99,dalessio_2013}, who found indications of a finite stability threshold in non-integrable systems as well.

To investigate this problem in a systematic way, we consider here a many-body analog of the Kapitza pendulum: the periodically-driven sine-Gordon model. This model is well suited for analytical treatments, including a high-frequency expansion, quadratic variational approaches, and renormalization-group methods. Unlike previously-studied spin systems, the present model has an unbounded single-particle energy spectrum \cite{note0}. Thanks to this property, infinite-temperature ensembles are characterized by an infinite energy density and are easily identified. We show the emergence of a sharp ``parametric resonance'', separating the absorbing (infinite temperature) from the non-absorbing (periodic) regimes. This transition survives in the thermodynamic limit and leads to a non-analytic behavior of the physical observables in the long time limit, as a function of the driving strength and/or frequency. We conjecture that this transition corresponds to a mean-field critical point of the many-body Floquet Hamiltonian. Our finding enriches the understanding of the coherent dynamics of parametrically forced system and paves the road toward the search of unconventional dynamical behavior of closed many-body systems.

\mysection{Review of a single Kapitza pendulum}

Before entering the domain of many-body physics, we briefly review the (well understood) case of a {\it classical} single
degree of freedom.
We consider the Hamiltonian of a periodically-driven simple (non-linear) pendulum, also known as Kapitza pendulum \cite{kapitza}, and described by the Hamiltonian
\be H(t) = \frac12p^2 - g(t)\cos(\phi),~~{\rm with} ~~g(t) =g_0+ g_1\cos(\g t)\;. \label{eq:HKapitza}\ee
Here $p$ and $\phi$ are canonically-conjugated coordinates satisfying $\lbrace p,\phi \rbrace =-{\textrm i}$, where $\lbrace\cdot,\cdot\rbrace$ are Poisson brackets. For $g_1=0$ the system displays two classical fixed points: a stable one at $\phi=0$ and an unstable one at $\phi=\pi$. (Throughout this paper we assume without loss of generality that $g_0>0$.).

In the presence of a periodic drive ($g_1\neq0$), the unstable fixed point can become dynamically stable. This counterintuitive result was first obtained by Kapitza \cite{kapitza} in the high-frequency limit $\gamma^2\gg g_0,g_1$. By averaging the classical equations of motion over the fast oscillations of the drive, Kapitza found that the ``upper'' extremum $\phi=\pi$ becomes stable for large enough driving amplitudes $g_1^2 > g_0 \gamma^2 /2$.  This pioneering work initiated the field of vibrational mechanics\cite{review_vibra}, and the Kapitza's method is used for description of periodic processes in atomic physics\cite{kapitza_meth_atoms}, plasma physics\cite{kapitza_meth_mechanics}, and cybernetical physics \cite{kapitza_meth_cyb}.

\begin{figure}[t]
\includegraphics[scale=0.25]{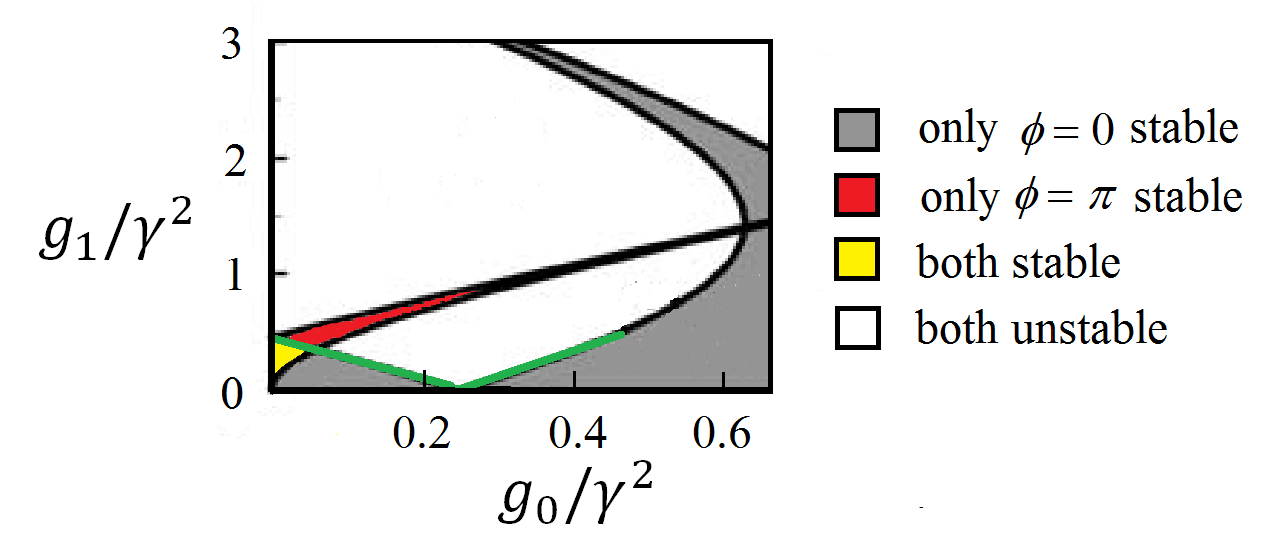}
\caption{Stability diagram of the classical Kapitza pendulum, adapted from Ref.~[\onlinecite{broer}]. In the colored areas at least one fixed point is stable, while in the white areas both minima are unstable and the system is fvariational methodsully ergodic. The parameters $g_0$, $g_1$ and $\gamma$ are defined in Eq.~(\ref{eq:HKapitza}) as $g(t)=g_0+g_1\cos(\gamma t)$. The green lines correspond to the stability threshold of the first parametric resonance, Eq. (\ref{eq:parametric})}\label{fig:broer}


\end{figure}

For finite driving frequencies $\gamma^2 \sim g_0,~g_1$ the lower fixed point ($\phi=0$) can become dynamically unstable as well. This transition can be analytically studied for example by applying the quadratic approximation $\cos(\phi)\to 1-\phi^2/2$ to Eq.~(\ref{eq:HKapitza}) (valid for small $g_1$). The resulting Hamiltonian corresponds to a periodically-driven harmonic oscillator, with frequency $\omega_0=\sqrt{g_0}$, driving frequency $\gamma$, and driving amplitude $g_1$. For infinitesimal driving frequencies ($g_1\to0$), this system displays parametric resonances at $\gamma=2\omega_0/n=2\sqrt{g_0}/n$, where $n$ is an integer\cite{LL}. For finite driving amplitudes each resonance extends to a finite region of driving frequencies\cite{LL}: in particular the first parametric resonance (n=1) extends to
\be 2 g_1 \le \left|\gamma^2 - 4g_0 \right|\; . \label{eq:parametric}\ee
Because the subsequent parametric resonances occur at lower frequencies, one finds that the point $\phi=0$ is always dynamically stable for driving frequencies that are larger than the critical value $\gamma_c=\sqrt{4g_0+2g_1}$, while regions of stability and instability alternate for $\gamma<\gamma_c$. In particular, the system is stable for $g_0/\gamma^2>0.25$ and small driving amplitudes. In what follows we will refer to these two stability regions as ``large driving frequency'' and ''large $g_0$'' respectively. As a side remark, we note that this quadratic approximation is valid for the quantum case as well, provided that the initial state is close to the $|\phi=0\langle$ state. In this case, the resulting stability phase diagram is expected to be the same.

Subsequent numerical studies of the classical Hamiltonian (\ref{eq:HKapitza}) lead to the stability diagram reproduced in Fig.~\ref{fig:broer} \cite{broer}. In this plot, the white areas represent regions in the parameter space in which both extrema are unstable and the system flows towards an infinite-temperature state, independently on the initial conditions. In contrast, in the colored regions at least one of the two extrema is stable, and the system is not ergodic. In this case the pendulum can be confined to move close to one of the extrema and will not in general reach a steady state described by an effective infinite temperature. The stability of the lower fixed point (green line in Fig.~\ref{fig:broer}) is well approximated by the boundaries of Eq.~(\ref{eq:parametric}). In particular, for $g_0=0$ (where the upper and lower extrema are mathematically equivalent), the system is dynamically stable for $g_1<g_c\approx 0.45\gamma^2$, and dynamically unstable for $g_1>g_c$. As a main result of this paper we will show that this transition remains sharply defined even in the many-body case.

\begin{figure}[t]
(a)\\
\includegraphics[scale=0.3]{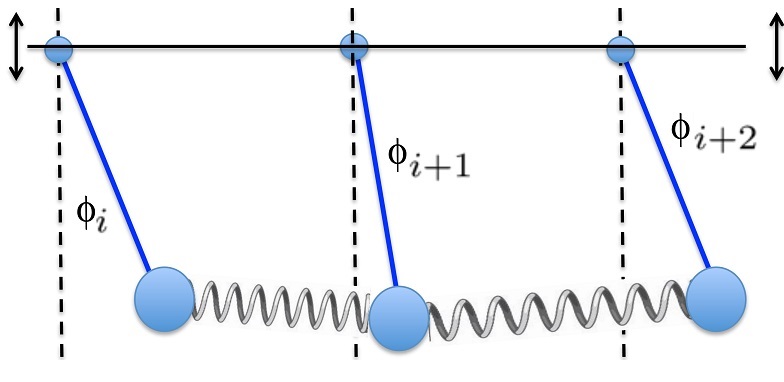}\\
(b)\\
\includegraphics[scale=0.8]{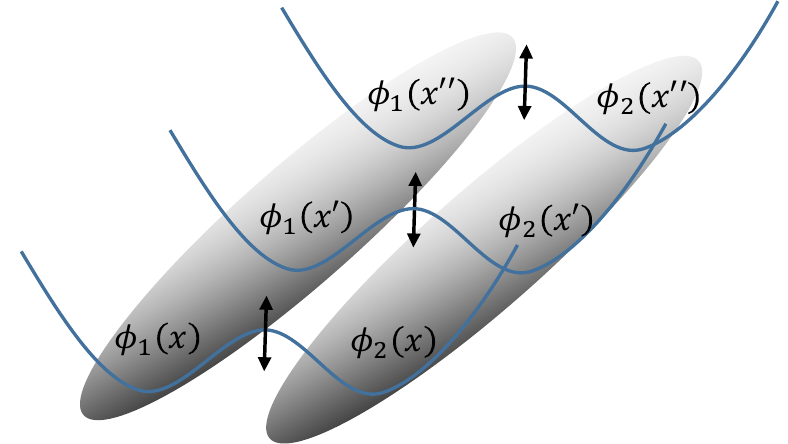}
\caption{Proposed physical realizations. (a) A one-dimensional array of Kapitza pendula, which are coupled and synchronously-driven realizes the time-dependent Frenkel-Kontorova model (\ref{eq:H_FK}). (b) Two coupled one-dimensional condensates, whose tunneling amplitude is periodically driven in time, realize the time-dependent sine-Gordon model (\ref{eq:H_in}). More details about the experimental realizations are presented in Sec.~\ref{sec:experiment}.}
\label{fig:discrete_model}
\end{figure}

\mysection{Many-body Kapitza pendulum}
To explore the fate of the dynamical instability in a many-body condition we consider an infinite number of coupled identical Kapitza pendula, depicted in Fig.~\ref{fig:discrete_model}(a). This system is described by the periodically-driven Frenkel-Kontorova \cite{braun98} model
\be H  = \Lambda\sum_i \left[\frac{K}2 P_i^2-\frac{1}{K} \cos(\phi_{i}-\phi_{i+1}) - \frac{g(t)}{\Lambda^2}\cos(\phi_i)\right]\;,\label{eq:H_FK}
\ee
where $P_i,~\phi_i$ are unitless variables satisfying $\lbrace P_j,~\phi_k\rbrace=-{\textrm i}\delta_{j,k}$, and $g(t)$ is defined in Eq.~(\ref{eq:HKapitza}). The energy scale $\Lambda$ determines the relative importance of the coupling between the pendula and the forces acting on each individual pendulum: in the limit $\Lambda\to0$ we expect to recover the case of an isolated periodically-driven pendulum. In the continuum limit, the model (\ref{eq:H_FK}) can be mapped to the periodically-driven sine-Gordon model
\be H = \int dx~\left[\frac{K}2 P^2+\frac{1}{2K} (\partial_x\phi)^2 - g(t)\cos(\phi)\right]\;,\label{eq:H_in}
\ee
$P(x)$ and $\phi(x)$ are canonically-conjugate fields,  $\lbrace P(x),\phi(x')\rbrace=-{\textrm i}\delta (x-x')$ and $K$ is the Luttinger parameter (we work in units for which the sound velocity is $u=1$). The parameter $\Lambda$ enters as an ultraviolet cutoff, setting the maximal allowed momentum: $\phi(x) = \int_{-\Lambda}^\Lambda dq/(2\pi)~e^{i q x} \phi_q$. The model (\ref{eq:H_in}) can also be realized using ultracold atoms constrained to cigar-shaped traps. In this case, the field $\phi=\phi_1-\phi_2$ represent the phase difference between the two condensates, and the time-dependent drive can be introduced by periodically modulating the transversal confining potential, as shown in Fig.~\ref{fig:discrete_model}(b). Because the fields $P$ and $\phi$ are continuous variable, the energy densities of the Hamiltonians (\ref{eq:H_FK}) and (\ref{eq:H_in}) are unbounded from above and allows to easily distinguish an absorbing behavior (in which the  energy density grows indefinitely in time) from a periodic one. This situation differs from previously-considered spin models, whose energy density is generically bounded from above.

\mysection{Infinite frequency expansion}
To understand the effect of periodic drives on dynamical instabilities and localization, it is convenient to define the effective Hamiltonian (often termed ``Floquet Hamiltonian'') $H_{\rm eff}$ as
\be U(T) = e^{-i H_{\rm eff} T}\;, \ee
where $U(T)$ is the evolution operator over one period of time $T=2\pi/\gamma$. In the case of the harmonic oscillator, it has been shown that the parametric resonance can be easily understood in terms of the effective Hamiltonian \cite{weigert}: in the stable regime the eigenmodes of $H_{\rm eff}$ are normalizable and any initial state can be expanded in this basis, leading to a periodic dynamics. In contrast, in the unstable regime  the eigenmodes of $H_{\rm eff}$ become not normalizable, in analogy to equilibrium Hamiltonians that are not bounded from below (such as $H=x^2-p^2$), and the dynamics becomes absorbing.

The Magnus expansion \cite{magnus_1954,blanes_2009} is an analytical tool to derive the effective Hamiltonian in the limit of large
drive frequency. The first-order term of the Magnus expansion, $H_{\rm eff}=(1/T)\int_0^T dt~ H(t)$, has a simple physical interpretation: when the driving frequency is infinite, the system perceives only the time-averaged value of $H(t)$. This observation has been successfully employed, for example, to engineer optical lattices with negative tunneling amplitude \cite{struck_2011,struck_2012,parker_2013}. In our case, the average Hamiltonian is simply described by the time-independent sine-Gordon model (Eq. (\ref{eq:H_in}) with $g_1=0$) and the drive has no effect on the system. The third-order Magnus expansion delivers:
\be H_{\rm eff} =H_{LL} + H' +H''- \int \frac{dx}{2\pi}~ \lbrack g_0\cos(\phi) + \tilde g\cos(2\phi)\rbrack\label{eq:Heff}\;,
\ee
where $H_{LL}$ is the Luttinger liquid Hamiltonian (Eq.~(\ref{eq:H_in}) with $g_1=g_0=0$),
\begin{align}
H'&=-g' \int dx P^2 \cos(\phi),\\
H''&=-g'' \int dx  (\partial_x \phi)^2 \cos (\phi);
\end{align}
$g' =\frac{g_1}{\gamma^2} K^2$, $g'' =\frac{g_1}{\gamma^2}$ and $\tilde g = K\frac{g_1}{\gamma^2}\left(\frac14g_1-g_0\right)$. See Appendix \ref{sec:magnus} for the details of this derivation. Eq.~(\ref{eq:Heff}) is analogous to the effective Hamiltonian of the single Kapitza pendulum \cite{dalessio_2013}.
The stability of the ``upper'' extremum $\phi=\pi$ is captured by the interplay between $g_0\cos(\phi)$ and $\tilde g\cos(2\phi)$: this point becomes dynamically stable when $g_0 < 4\tilde g$, or equivalently when $g_0<\frac{g_1^2K}{\gamma^2}$, taking
into account that the Magnus expansion is valid in the limit\cite{note} of $g_0/\gamma^2 \ll 1$.

The term $H'$ leads to the dynamical instability of the system: having a negative sign, it suppresses the kinetic energy $\sim P^2$,
eventually leading to an inversion of its sign. Using a quadratic variational approach, we can approximate
$H'\approx -4g'\int dx \av{\cos(\phi)}P^2$.
The stability of the $\phi=0$ minimum can then be studied through $\av{\cos(\phi)}\approx 1$, leading to a renormalized kinetic energy $K P^2/2 - K^2 (g_1/\gamma^2) P^2 = K_{\rm eff} P^2/2$, with
\be K_{\rm eff} = K \left(1 - 2 K\frac{g_1}{\gamma^2}\right) \label{eq:Keff}\ee
This expression indicates that the system is dynamically stable for $K g_1 /\gamma^2 < 0.5$. For larger driving amplitudes (or smaller driving frequencies) the kinetic energy becomes negative and the system becomes unstable.

Higher-order terms in the Magnus expansion can be used to determine the qualitative dependence of the critical driving amplitude as a function of $\Lambda/\gamma$ and $K g_1/\gamma^2$. For example, the fifth-order term of the Magnus expansion contains terms proportional to $g_1(t)/\gamma^4 \lbrace(\partial_x \phi)^2,\lbrace(\partial_x \phi)^2,\lbrace P^2,\lbrace P^2,\cos \phi\rbrace\rbrace\rbrace\rbrace \sim P^2 \partial^2_x\cos(\phi)$, which renormalizes $g'$ by a factor of $ 8\frac{g_1^2}{\gamma^2}\left( \frac{\Lambda}{\gamma}\right)^2$. This positive contribution leads to a decrease of the critical driving amplitude as a function of $\Lambda/\gamma$. As mentioned above (see Eq.(\ref{eq:H_FK})), $\Lambda$ sets the coupling between the Kapitza pendula and is indeed expected to shrink the stability region of the system.

Since the Magnus expansion generates an infinite number of terms, one may suspect that the full series could renormalize the critical amplitude to zero (making the system always dynamically unstable). To address this point, we now resort to the quantum version of the problem, where powerful analytical techniques are available.

\section{Renormalization group arguments}

Let us now consider the quantum version of the Hamiltonian (\ref{eq:H_in}). In the absence of a drive ($g_1=0$), this model corresponds to the celebrated sine-Gordon model\cite{giamarchi_book,gogolin_book}. Its ground state displays a quantum phase transition of the Kosterlitz-Thouless type: for $K>K_c=8\pi - o(g/\Lambda)$ the cosine term is irrelevant and the model supports gapless excitations, while for $K<K_c$ the system has a finite excitation gap $\Delta$. The presence of a gap is known to have significant effects on the response of the system to low-frequency modes, by exponentially suppressing the energy absorption. In contrast, at large driving frequencies the excitation gap is expected to have little effect.

To analyze the limit of large driving frequencies, we propose to consider the Floquet Hamiltonian, as defined by the Magnus expansion \footnote{The eigenvalues of the Floquet Hamiltonian are defined only up to $\gamma$, but the Magnus expansion implicitly specifies one particular choice}. Specifically, we consider the ground-state properties of this Hamiltonian, which can be conveniently studied through the renormalization group (RG) approach. The existence of a well-defined ground state for the Floquet Hamiltonian implicitly demonstrates the ergodicity of the system: unstable systems are generically expected to have non-normalizable eigenstates \cite{weigert}.

In our case, the first-order term of the Magnus expansion corresponds to the well-known sine-Gordon model. When $K_{\rm eff}>8\pi$ this Hamiltonian flows under RG towards the Luttinger liquid theory (Eq.(\ref{eq:H_in}) with $g(t)=0$). Higher order terms are given by commutators of the Hamiltonian at different times (see Appendix \ref{sec:magnus}). Because the time-dependent part of our Hamiltonian is proportional to $\cos(\phi)$, each commutator necessarily includes $\cos(\phi)$, or its derivatives. With respect to the Luttinger-liquid fixed point, these terms are irrelevant in an RG sense, and are not expected to affect the ground-state-properties of the Floquet Hamiltonian. If this is indeed the case, the (asymptotic) expectation values of physical operators such as $|\phi_q|^2$ and $|P_q|^2$ are finite (and proportional to $K_{\rm eff}/|q|$ and $|q|/K_{\rm eff}$ respectively), indicating that the system does not always flow to an infinite-temperature ensemble. When the frequency is reduced, the amplitude of higher-order terms of the Magnus expansion increases: although irrelevant in an RG sense, if sufficiently large, these terms can lead to a transition towards an unstable regime.

Many-body quantum fluctuations have an important effect on the stability of the inverted pendulum $\phi=\pi$ as well. As mentioned above, this effect is related to the interplay between $\cos(\phi)$ and $\cos(2\phi)$ in Eq.~\ref{eq:Heff}. In the ground-state of this Hamiltonian, quantum fluctuations change the scaling dimension of an operator $\cos(\alpha\phi)$ to $2-\alpha^2 K/4$.
Thus for finite $K_{\rm eff}$, the term with $\tilde g$ is less relevant than the term with $g_0$
from a renormalization group (RG) point of view, making the upper extremum less stable than in the case of a simple pendulum.
A simple scaling analysis reveals that the stability boundary between the two extrema at $0$ and $\pi$ is renormalized
by approximately $(\tilde g/\Lambda)^2$, where $\Lambda$ is the theory cutoff. This is a strong indication that, for any finite $\Lambda$,
the extremum at $\pi$  can still be made stable with large enough driving amplitudes\cite{note1}.

\begin{figure}[b]
\centering
\includegraphics[scale=0.6]{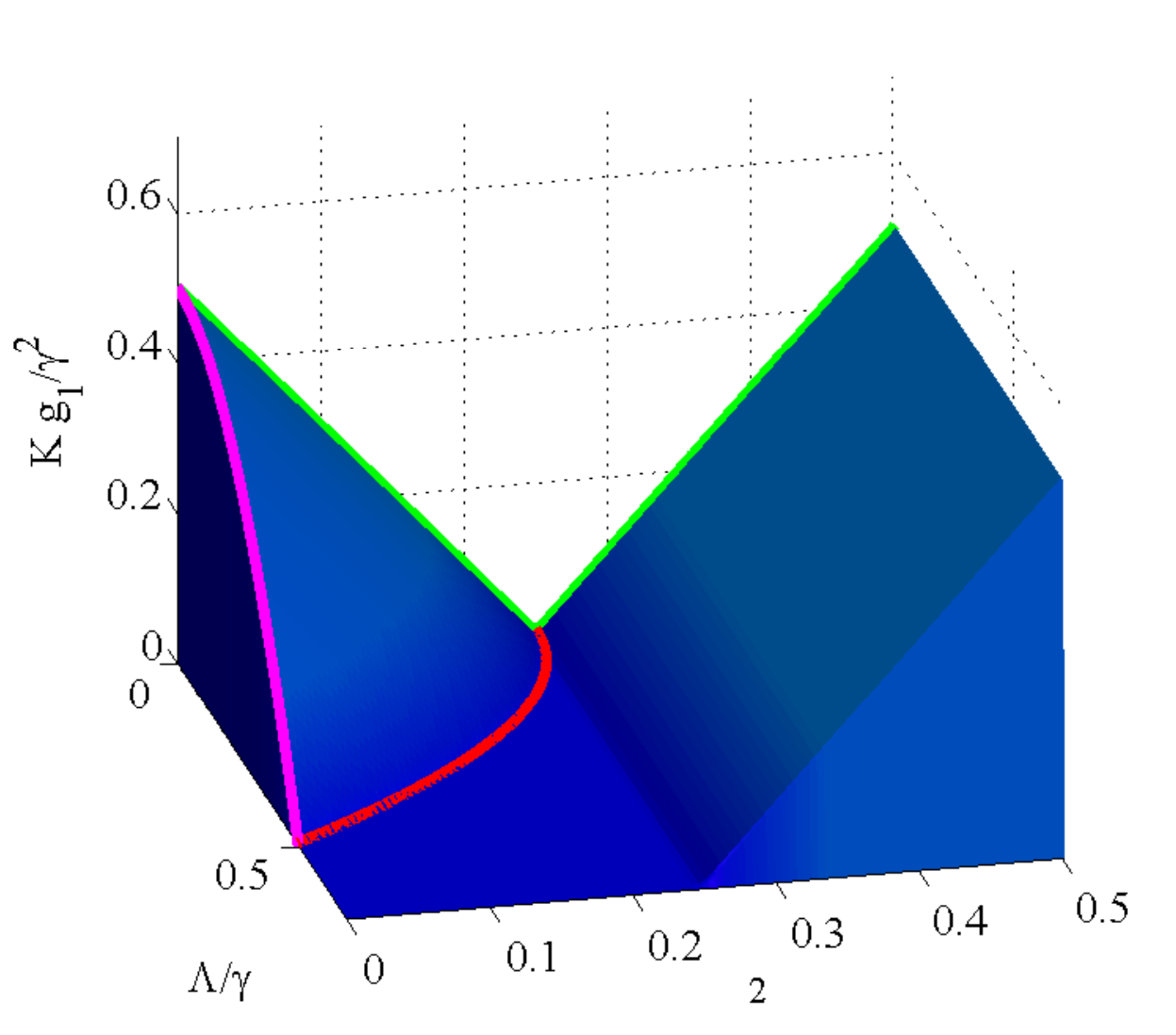}
\caption{Stability diagram of the periodically-driven sine-Gordon model, as obtained from the quadratic expansion of the lower minimum. The system displays two distinct stability regions at large $\gamma$ and large $g_0$, respectively (see text). The green line refers to the limit $\Lambda/\gamma\to0$, where the problem maps to the stability of the lower minimum of the Kapitza pendulum (green line of Fig.~\ref{fig:broer}).}\label{fig:phasediagram}
\end{figure}

\section{Quadratic expansion}
In analogy to the single Kapitza pendulum, the simplest way to tackle the many-body Hamiltonian (\ref{eq:H_in}) is to expand the cosine term to quadratic order. In particular, to analyze the stability of the $\phi=0$ minimum we can use $\cos(\phi)\to 1-\phi^2/2$. In this approximation the system becomes equivalent to a set of decoupled harmonic oscillators with Hamiltonian $H=\sum_q H_q$, where
\be H_q  = \frac{K}2 P_q^2+ \frac1{2K}\left(q^2 + K g_0 + K g_1\cos(\gamma t)\right)\phi_q^2\;.\label{eq:H_quad}\ee
Eq.~(\ref{eq:H_quad}) corresponds to the Hamiltonian of a periodically driven harmonic oscillator, whose stability diagram is well known. In the limit of $g_1\to0$, the system is dynamically stable as long as $\gamma> 2\max [\omega_{q}] =2\max[\sqrt{q^2+K g_0}] = 2\sqrt{\Lambda^2+K g_0}$, or  $\gamma< 2\min [\omega_{q}] =2 \min [\sqrt{q^2+K g_0}] = 2\sqrt{K g_0}$. In analogy to Eq.(\ref{eq:parametric}), for finite $g_1$ the stability condition is modified to
\be 2 K g_1 < \max\left[\gamma^2 - 4\left(K g_0 + \Lambda^2\right),  4\left(K g_0\right) - \gamma^2\right] \;\label{eq:trans_quad}\; \ee
The resulting dynamical phase diagram is plotted in Fig.~\ref{fig:phasediagram}. In the limit of $\Lambda\to0$ we recover the stability diagram of the lower minimum of an isolated Kapitza pendulum: this is demonstrated by the quantitative agreement between the green curves of Fig.\ref{fig:broer} and Fig.~\ref{fig:phasediagram}.

For finite $\Lambda$ we observe two distinct stability regions, characterized by a different dependence on the ultraviolet cutoff $\Lambda$. The first stability region is adiabatically connected to the limit of infinite driving-frequency, $\gamma\to\infty$ (the origin in Fig.~\ref{fig:phasediagram})). Its boundaries are described by
${K g_1}/2\gamma^2 + {K g_0}/\gamma^2 +{\Lambda ^2}/{\gamma^2} = 0.25$. For $g_0\to0$ and $\Lambda\to0$ this expression precisely coincides with the result obtained from the large-frequency Magnus expansion, $K g_1/\gamma^2=0.5$. The stability region is strongly suppressed by $\Lambda$ and eventually disappears at $\Lambda/\gamma \approx 0.5$. This indicates that it is related to the stability of the degrees of freedom at the shortest length scales $\Lambda^{-1}$: its character is therefore predicted to be analogous to the stability of a single Kapitza pendulum. In contrast, the second stability region, at large $g_0$ is roughly independent on $\Lambda/\gamma$. This stability region is determined by the resonant excitation of the lowest frequency collective excitation of the system in the static ($\gamma \rightarrow 0$) regime, where the system is dynamically stable due to the presence of a gap, $\Delta$. In the quadratic approximation, the gap is given by $\Delta\approx \sqrt{K g_0}$, and the stability condition $\gamma<2\Delta$ reads $K g_0 /\gamma^2 < 0.25$.

The present quadratic expansion bares a close resemblance to the analysis of Pielawa \cite{pielawa_2011}, who considered periodic modulations of the sound velocity (but keeping $g_0=0$). The validity of these quadratic approximations is however undermined by the role of non-linear terms in non-equilibrium situations (and specifically noisy environments \cite{dallatorre_2012} and quantum quenches \cite{mitra_2012}). Even if irrelevant at equilibrium, non-linear terms enable the transfer of energy between modes with different momentum and are therefore necessary to describe thermalization. The question which we will address here is whether mode-coupling effects are sufficient to destroy the dynamical instability described above.

\mysection{Self-consistent variational approach}

The above-mentioned quadratic approximation can be improved by considering a generic time-dependent Gaussian wavefunction, along the lines of Jackiw and Kerman \cite{jackiw}:
\be
\Psi_v[\phi(x)] = \mathcal{A}\exp\left(-\int_{x,y}\,\phi(x) \left[\frac14G^{-1}_{x,y} - i\Sigma_{x,y}\right]\phi(y)\right)\label{eq:var} \ee
Eq.~(\ref{eq:var}) is a particular case of the generic wavefunction proposed by Cooper \etal\cite{Cooper2003}, valid when the expectation value of the field and its conjugate momenta are zero: $\langle \phi(x)\rangle =\langle P(x)\rangle=0$.
The operator $\mathcal{A} \sim (\mathrm{det}\,G)^{1/4}$ is the normalization constant to ensure the unitarity of the evolution at all times:
\begin{equation}
\langle \Psi_v| \Psi_v \rangle = \int \mathcal{D}[\phi]\Psi^*_v[\phi]\,\Psi_v[\phi] = 1.
\end{equation}

The functions $G_{x,y}$ and $\Sigma_{x,y}$ are variational parameter to be determined  self-consistently. To this end, we invoke the Dirac-Frenkel variational principle and define an effective classical Lagrangian density as follows:
\begin{equation}
\mathcal{L}_{\rm eff} = \int \mathcal{D}[\phi]\,\Psi^*_v[\phi]\left(i\partial_t - \mathcal{H}[\phi,\partial/\partial \phi]\right)\Psi_v[\phi]
\end{equation}
For a translation-invariant system, we find $G_{x,y} \equiv G_{x-y} \equiv 1/(2\pi)\int_{-\Lambda}^\Lambda \dd k\,G_k\,e^{ik(x-y)}$, where $\Lambda$ is the UV cutoff, and the effective action $S_{\rm eff} \equiv \int \dd t\int \dd x\,\mathcal{L}_{\rm eff}$ is given by
\begin{align}
S_\CL &=  Z(t)\,g(t) + \int_{-\Lambda}^\Lambda\frac{\dd k}{2\pi} \Big(\, \Sigma_k \dot{G}_k \nonumber \\
&-\frac{1}{8} \, K \, G^{-1}_k - 2 K \,\Sigma_k G_k \Sigma_k - \frac{k^2}{2 K}\,G_k\Big),\nonumber\\
{\rm where} ~~~ & Z(t) = \exp\left(-\frac{1}{2}\int_{-\Lambda}^\Lambda \frac{\dd k}{2\pi} \, G_k\right). \label{eq:action}
\end{align}

The equations of motion are given by the saddle point of the effective action. By requiring $\delta S_{\rm eff}/\delta G=\delta S_{\rm eff}/\delta \Sigma=0$ we obtain
\begin{subequations}\label{eq:EOM}
\begin{align}
\dot{G}_k &= 4K\,G_k\Sigma_k,\\
\label{eq:sigdot}
\dot{\Sigma}_k &= \frac{1}{8}\,K G^{-2}_k - 2K\,\Sigma_k^2 -\frac{k^2}{2K} - \frac{1}{2}\,Z(t)\,g(t)\,
\end{align}
\end{subequations}

In the following calculations we assume the system to be initially found in a stationary state satisfying
\begin{equation}
G_k = \frac{K}2\,\frac{1}{\sqrt{k^2 + \Delta_0^2}},
\end{equation}
where $\Delta_0$ is self-consistently given as:
\begin{equation}
\Delta_0^2 = g_0 K\,\exp\left(-\frac{1}{2}\int_{-\Lambda}^\Lambda \frac{d k}{2\pi}\frac K2\,\frac{1}{\sqrt{k^2 + \Delta_0^2}}\right).
\end{equation}
Assuming $\Delta_0 \ll \Lambda$, the above equation gives $\Delta_0^2 \approx (g_0 K/2)[\Delta_0/(2\Lambda)]^{K/4\pi}$, which implies a critical point at $K_c =8\pi$ (Kosterlitz-Thouless transition). The cosine is relevant (irrelevant) for $K<K_c$ $(K>K_c)$.

\begin{figure}[b]
\center
(a) $K=0.1\pi$\\
\includegraphics[scale=0.2]{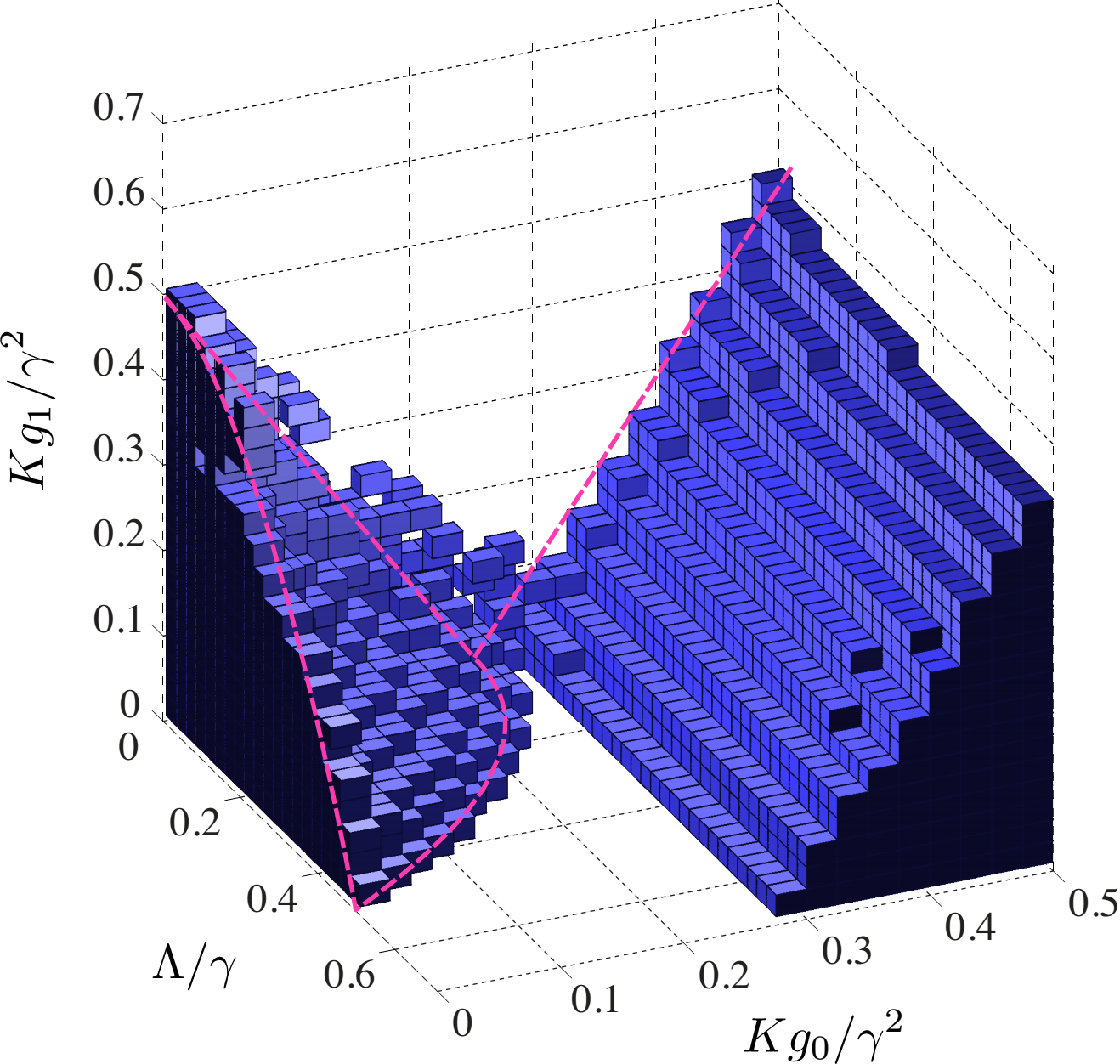}\\\vspace{0.25cm}
(b) $K g_0 /\gamma^2 = 10^{-4}$\\
\includegraphics[scale=0.2]{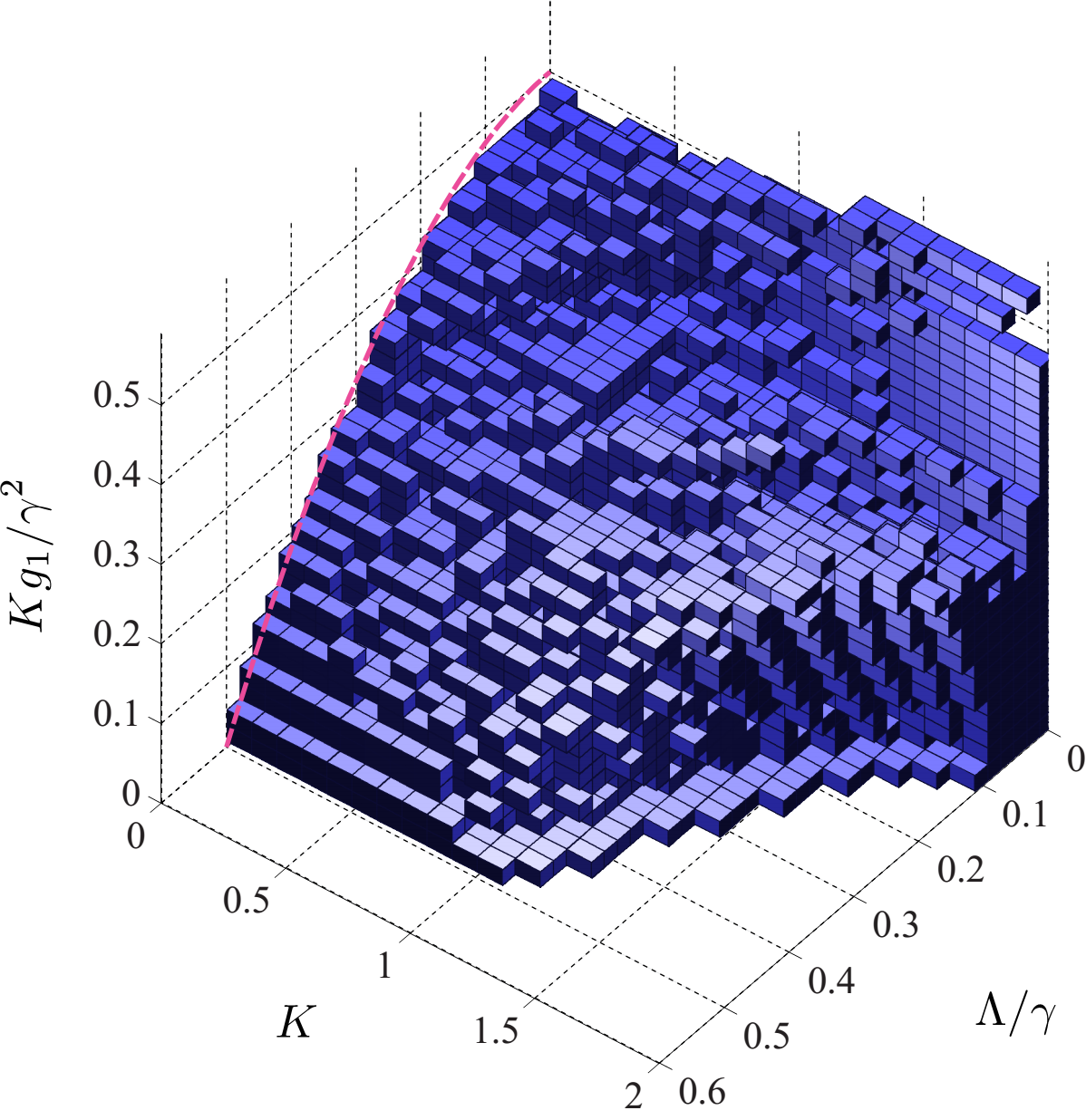}
\caption{The dynamical regimes (a) as a function of $(g_0 K / \gamma^2, g_1 K / \gamma^2, \Lambda/\gamma)$ for fixed $K=0.1$, (b) as a function of $(K, g_1 K / \gamma^2, \Lambda/\gamma)$ for fixed $g_0 K / \gamma^2 = 10^{-4}$. Blue (white) regions correspond to points in the parameter space where the system is stable (unstable). Solid lines in (a) correspond to the results of the quadratic approximation (Fig.~\ref{fig:phasediagram}) and identify two distinct stability regions, respectively at large $\gamma$ and large $g_0$. In this plot the stability criterion is arbitrarily set to $Z(T_f)/Z(0) > 0.95$ with $T_f = 100 (2\pi/\gamma)$ (See also Fig.~\ref{fig:mehrtash2} for details about the finite-time scaling.)}
\label{fig:mehrtash}
\end{figure}

In this initial gapped phase, the classical oscillation frequency, $\sqrt{K g_0}$ is renormalized by the factor $Z$ due to quantum fluctuations (see Eqs.~(\ref{eq:action}) and~(\ref{eq:EOM})). The introduction of a modulation to the amplitude of the bare cosine potential leads to one of the two following scenarios (i) Unstable (ergodic) regime: the driving field amplifies quantum fluctuations (i.e. leads to ``particle generation'' via parametric resonance) and closes the gap, i.e. $Z(t) \rightarrow 0$. Once the gap closes, it remains closed; we take this as an indication of the runaway to the infinite-temperature limit (we can also study the absorbed kinetic energy in this formalism as well); (ii)  Stable (non-ergodic) regime:  quantum fluctuations remain bounded, $Z(t)$ stays finite at all times, and $\phi$ remains localized. These two regimes are indicated in Fig.~\ref{fig:mehrtash} as white and steel-blue regions, and are in quantitative agreement with the results of the quadratic approximation, Fig.~\ref{fig:phasediagram}. The apparent inconsistency for small $\Lambda/\gamma\ll1$ and $K g_1/\gamma^2\ll1$ is simply due to finite-time effects (see also Appendix~\ref{app:single})

The effects of quantum fluctuations is analyzed in Fig.~\ref{fig:mehrtash}(b). This plot displays the stability  diagram for a fixed and small $g_0$ ($g_0 K/\gamma^2 = 10^{-4}$) as a function of $g_1$, $K$ and $\Lambda$. For small $K$ we reproduce the large-frequency stability lobe of Fig.~\ref{fig:mehrtash}(a). Indeed, the quadratic approximation is expected to become exact in the limit of $K\to0$. Finite values of the Luttinger parameter $K$ significantly shrink the volume of the non-ergodic (stable) regime. This result indicates that many-body quantum fluctuations promote ergodicity. At the same time, our analysis indicates that a finite region of stability can survive in the thermodynamic limit, even in the presence of quantum fluctuations.

\begin{figure}[b]
\centering
\includegraphics[scale=0.5]{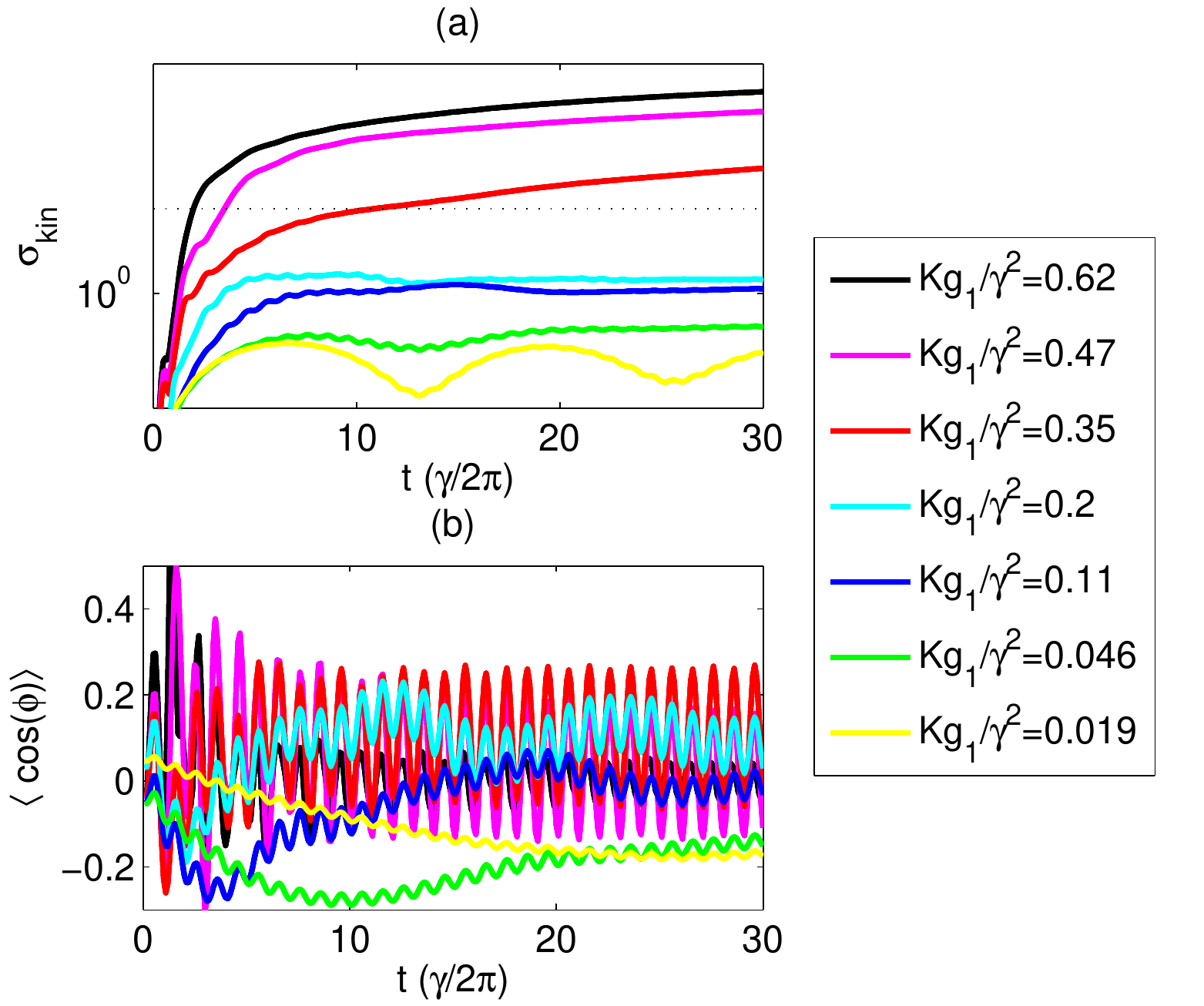}
\caption{(a) Time evolution of the normalized kinetic energy
$\sigma_{\rm kin}(t)=E_{\rm kin}(t)/E_{\rm kin}(t=0)-1$, with
$E_{\rm kin}(t)=(1/K)\langle (\partial_x\phi)^2 \rangle$ for $K=0.4\pi$, $g_0=0$, $\Lambda/\gamma=0.04$, $L=200$, $N=400$. For small drives and large frequencies (lower curves) the system is stable and periodically oscillates with a period $\pi/\Lambda = 12.5 (2\pi/\gamma)$. Upon reaching a critical value of $K g_1/\gamma^2$ the oscillations are substituted by an exponential increase of the energy. The dashed line is a guide for the eye. (b) Time evolution of $\langle \cos(2\phi)\rangle $ for the same parameters as before. The oscillations become very large around a critical value of $K g_1/\gamma^2$.}\label{fig:figure3AB}
\end{figure}

\begin{figure}[t]
\centering
\includegraphics[scale=0.5]{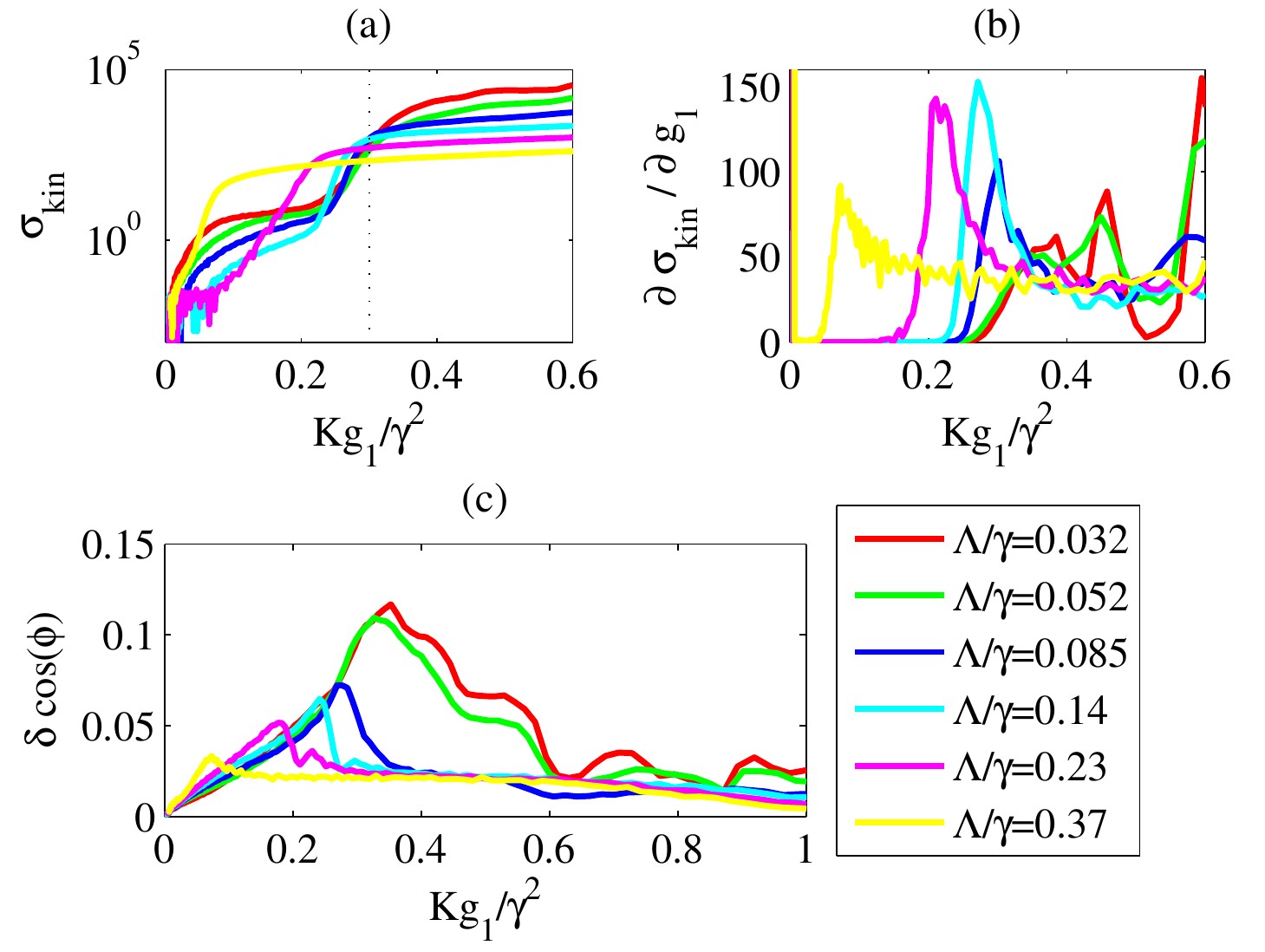}
\caption{(a) Normalized kinetic energy $\sigma_{\rm kin}$ at long times ($t =100\times 2\pi/\gamma$) as a function of the normalized driving amplitude $K g_1/\gamma^2$ for $K=0.4\pi$, $g_0=0$, $L=200$. This quantity displays a sharp kink around a critical value indicated by the dashed line. (b) First derivative of the kinetic energy, showing a sharp discontinuity in its derivative, confirming the hypothesis of a second-order-like phase transition. (b) Oscillation amplitude of $\av{\cos(2\phi)}$ at long times ($T=30\times 2\pi/\gamma$) as a function of the normalized driving amplitude $K g_1/\gamma^2$.
This quantity displays a sharp peak around a critical value, corresponding to the sharp increase in the kinetic energy identified in (a). With increasing $\Lambda/\gamma$ the peak moves to lower values of the drive amplitude and becomes less pronounced.}
\label{fig:figure4ABC}
\end{figure}

\mysection{Semiclassical dynamics}
\label{sec:TWA}
To further demonstrate the existence of an instability transition at a finite driving frequency, we now numerically solve the classical equations of motion associated with the Hamiltonian~(\ref{eq:H_in}). Specifically, we focus here on the stability region at large driving frequencies and along $g_0=0$ (magenta curve of Fig.s~\ref{fig:phasediagram} and \ref{fig:mehrtash}(a)). Following the truncated-Wigner prescription \cite{polkovnikov_2009,lancaster}, we randomly select the initial conditions from a Gaussian ensemble corresponding to the ground state of (\ref{eq:H_in}) with $g(t)=0$ and solve the classical equations of motion associated with the Hamiltonian (\ref{eq:H_in}). Although not shown here, we checked that other choices of initial conditions lead to qualitatively similar results. Fig.~\ref{fig:figure3AB} shows (a) the time evolution of the average kinetic energy and (b) of the expectation value $\av{\cos(\phi)}$ for different driving amplitudes. The former displays a sharp increase in correspondence of the expected dynamical transition. Fig.~\ref{fig:figure4ABC}(a) shows the average kinetic energy $E_{\infty}$, and the oscillation amplitude $\delta\cos(\phi)$ as a function of driving frequency: Although $E_{\infty}$ is smooth as a function of the driving amplitude, its first derivative (Fig.~\ref{fig:figure4ABC}(b)) shows a sharp kink at a critical value of $K g_1/\gamma^2$. Non-discontinuities in the second derivative of the energy are clear evidence of continuous transitions. In contrast, $\delta\cos(\phi)$ presents
a kink itself at the critical value of $K g_1/\gamma^2$.

\begin{figure}[t]
\centering
\includegraphics[scale=0.6]{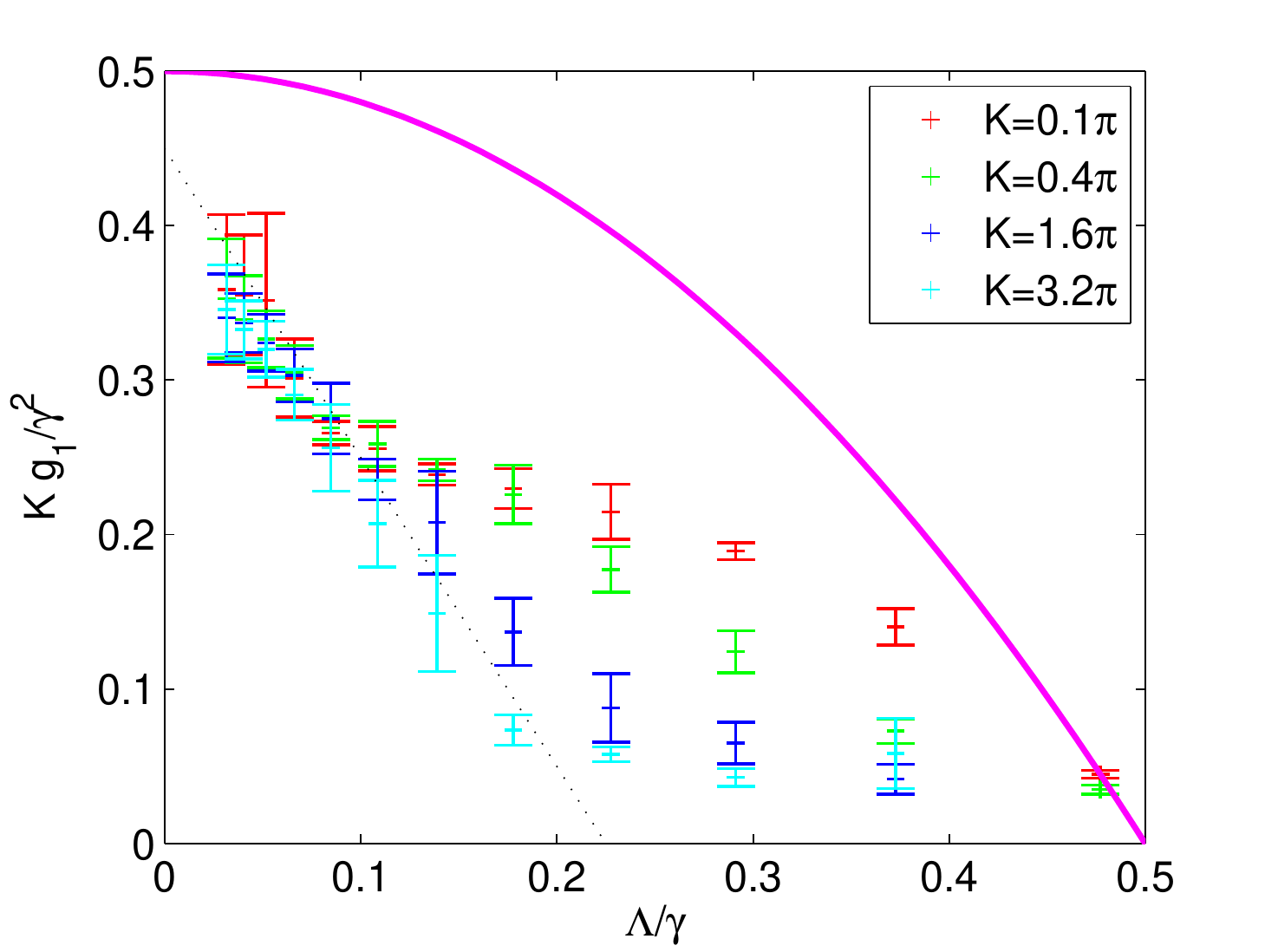} \\
\caption{Stability diagram: critical drive as a function of the UV cutoff $\Lambda/\gamma$ for $g_0=0$, as obtained through the present semiclassical approximation (+). The dashed line is a guide for the eyes. In the limit $\Lambda/\gamma\to0$ all the curves tend towards the critical value of a single Kapitza pendulum, $K g_1/\gamma^2\approx0.45$. The error bars refer to the numerical uncertainty in the position of the peak and demonstrate that the peaks do not broaden and remain well defined  for finite $\Lambda/\gamma$. The magenta curve corresponds to the result of the quadratic expansion (Eq.(\ref{eq:trans_quad}) and magenta curve in Fig.~\ref{fig:phasediagram}).}\label{fig:figure5}
\end{figure}

From the position of the kink in either Fig.~\ref{fig:figure4ABC}(b) or (c) we compute the dynamical phase diagram shown in Fig.~\ref{fig:figure5}. We find that larger $K$ lead to a reduction of the stability region. A similar result was obtained using the self-consistent variational approach (Fig.~\ref{fig:mehrtash}). In particular, the stability diagram is mainly insensitive to the initial conditions. Specifically, we repeated the semiclassical dynamics for a different set of initial conditions (corresponding to an initially gapped state with local correlations only) and observed a qualitatively similar dynamical phase diagram.

\begin{figure}[t]
\vspace{-0.2cm}
\includegraphics[scale=0.8]{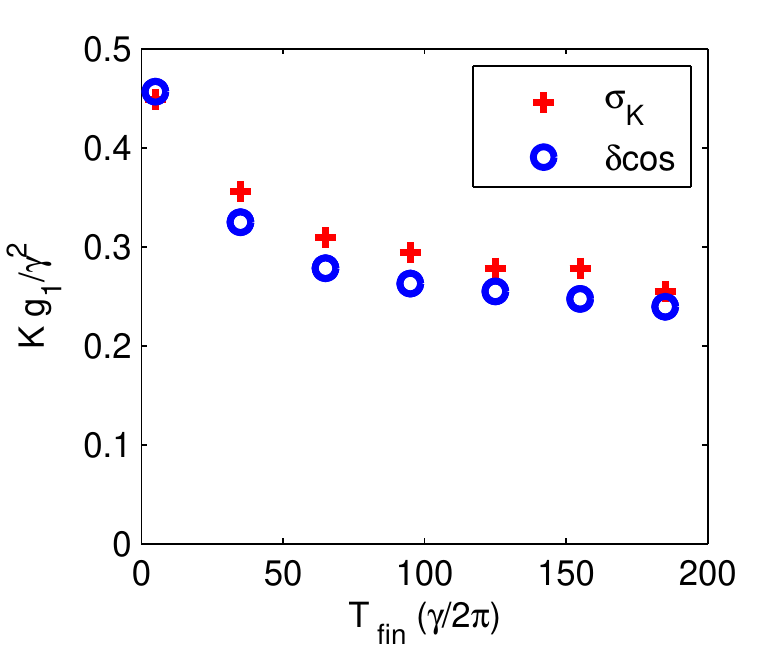}
\caption{Critical value of the driving field as a function of the waiting time $T_{\rm fin}$. The critical field is determined by two independent methods, namely by observing the peaks of $\sigma_K$ as in Fig.~\ref{fig:figure4ABC}(b), and of $\delta\cos$, as in Fig.~\ref{fig:figure4ABC}(c). Both methods show an initial decrease of the critical drive, followed by a saturation at a finite asymptotic value. Numerical values: $g_0=0$,~$K=0.4\pi$,~$\Lambda/\gamma=0.1$}\label{fig:asymptotic}
\end{figure}

The above calculations refer to physical observables measured after a finite number of driving periods. One important question regards the evolution of the phase diagram of Fig.~\ref{fig:phasediagram} as a function of  time. In particular, one may wonder whether the stability region gradually shrinks and disappears in the infinite-time limit. In other words, can the dynamical transition be induced by applying an infinitesimal drive for very long times? To address this question, in Fig.~\ref{fig:asymptotic} we plot the critical driving amplitude as a function of the number of driving periods. Although the critical driving amplitude initially decreases as a function of time, we observe that at long times it tends to a finite asymptotic value: the stable regime occupies a finite region in the parameter space even in the asymptotic long-time limit. The number of driving periods plays an analogous role to the size of the system in equilibrium phase transitions: when appropriately rescaled, any physical quantity is associated with a well-defined asymptotic scaling limit (see Appendix \ref{app:scaling}).

The present calculations demonstrate the existence of a localized (non-ergodic) phase in the thermodynamic limit. Following D'Alessio \etal \cite{dalessio_2013}, this result can be interpreted as a (many-body) energy localization transition. Remarkably, the present semiclassical approach involves the solution of classical equations of motion (subject to quantum initial conditions). Our findings are in contrast to an earlier conjecture formulated in the context of MBL systems with randomness by Oganesyan \etal \cite{oganesyan2009} and suggesting that such a transition has a pure quantum origin and does not occur in classical systems.

\mysection{Experimental Realization and Outlook}
\label{sec:experiment}
We now offer more details about the two physical realizations that were anticipated in Fig.~\ref{fig:discrete_model}. The two proposed experiments belong respectively to the classical and quantum realm. The present calculations suggest that the two model should have a qualitatively similar stability diagram, but an experimental verification is required. The realization of the Hamiltonian (\ref{eq:H_FK}) using classical elements seems particularly appealing and relatively easy: this model describes an array of pendula attached to a common periodically oscillating support, and coupled through nearest neighbor couplings (see Fig.\ref{fig:discrete_model}(a)).

A quantum version of the periodically-driven sine-Gordon model (\ref{eq:H_in}) can be realized using ultracold atoms\cite{schmiedmayer_2000,zimmermann_review,bouchoule_review,bouchoule_2003}. By trapping the atoms in cigar-shaped potentials it is possible to obtain systems in which the dynamics is effectively one dimensional. Specifically, the transversal confinement can be generated through laser standing waves (see for example Ref.~[\onlinecite{bloch2005}]), or through magnetic fields induced by currents running on a nearby chip (see for example Ref.~[\onlinecite{langen2014}]). In both cases, the amplitude of the transverse confinement can be easily modulated over time, allowing to realize the setup shown in Fig.~\ref{fig:discrete_model}(b). Here a time-dependent transverse confinement induces a time-dependent tunneling coupling between two parallel tubes. In model (\ref{eq:H_in}) this coupling is modeled as $\int dx\cos(\phi)$, where  $\phi(x)=\phi_1(x)-\phi_2(x)$ is the local phase difference between the
quasi-condensates\cite{atom_chip_inter,double_well_atom_chip_inter,gritsev_coupled_condensates,superfluid_junction,dallatorre_2014,coupled_ll_thierry}. Dealing with a system of two coupled tubes, the Luttinger parameter is doubled with respect to a single tube and equals to $K =2\sqrt{\Gamma}$, where $\Gamma= m \Lambda/\hbar^2\rho_0$, $\Lambda=\mu$ is the chemical potential, $m$ is the mass of the atoms, and $\rho_0$ their average one-dimensional density \footnote{With respect to the conventions of Ref.~[\onlinecite{cazalilla2011}] the Luttinger parameter of a single tube is first inverted to account from the transition between the phase and density representations, and then multiplied by $\pi$ due to the different choice of commutation relations.}. Atoms on a chip are characterized by relatively small interaction energies, constraining the maximal value of the achievable $K$. Typical experimental values of $\gamma$ are in the order of $\gamma\lesssim {10^{-2}}$, or $K\lesssim 0.2$. The parameter $g(t)$ is set by the instantaneous (single-particle) tunneling rate through $J_\perp(t) = K g(t)/\mu$. These experiments are therefore constrained to $g(t)>0$, or $g_1<g_0$. Fig.s \ref{fig:phasediagram} and \ref{fig:mehrtash} show that the a wide region of stability is expected in the physically relevant regime (small $K$ and $g_1<g_0$), demonstrating the feasibility of the proposed experiment.

An alternative procedure to periodically drive the two coupled tubes involves the modulation of the chemical potential difference between them $\delta\mu(t)=\mu_1(t)-\mu_2(t)$ (while keeping approximately fixed the tunneling element). In the bosonization language of Eq.(\ref{eq:H_in}), this corresponds to a time-dependent field that couples to the atomic density. An appropriate gauge transformation allows one to map this problem into a phase-modulated sine-Gordon model in which the drive enters through the phase drive $\phi_0(t)$ as $g\cos(\phi-\phi_0(t))$. We postpone the detailed analysis of this model to a future publication: preliminary calculations indicate that the stability diagram is analogous to Fig.~\ref{fig:phasediagram}. However, we do not know yet whether the two models belong to the same universality class.

The sine-Gordon model is often discussed in the context of isolated tubes under the effect of a longitudinal standing waves, or optical lattice, as well. At equilibrium this model describes the Luttinger liquid to Mott insulator quantum phase transition of one dimensional degenerate gases (see Ref.~\cite{cazalilla2011} for a review). Interestingly, the inverted pendulum $\phi=\pi$ corresponds to a distinct topological phase, the Haldane insulator~\cite{dallatorre_2006,berg_2008}. At equilibrium non-local interactions are needed in order to stabilize this phase. As explained above, in the presence of a periodic drive it may be possible to stabilize it by tuning the system in the region where only the ``upper'' extremum is stable. This situation is analogous to the recent proposal of Greschner \etal~\cite{Greschner2014}. In relation to the dynamical stability of this phase, one however needs to notice that the optical lattice affects the Luttinger parameter as well. A correct description of these experiments therefore involves the additional transformation  $K \to K(t) = K_0 + K_1\cos(\gamma t)$, with possible significant consequences for the resulting stability diagram. Finally, it is worth mentioning other possible realizations of the model (\ref{eq:H_in}), including for example RF coupled spinor condensates \cite{spinorcondensates} and arrays of Josephson junctions \cite{arraysofJJ} in the presence of time-dependent magnetic fields .

To summarize, in this paper we combined different analytical and numerical tools to study the periodically-driven sine-Gordon model. We applied a controlled high-frequency expansion, the Magnus expansion, to derive an effective (Floquet) Hamiltonian. Employing ideas from the renormalization group (RG) method, we propose  the existence of a non-absorbing (non-ergodic) fixed point, in which the system is weakly affected by the periodic drive. At a critical value of the driving frequency (or equivalently of the driving amplitude), the system undergoes a dynamical phase transition and flows towards the infinite temperature absorbing (ergodic) state. The transition occurs at a finite value of the drive amplitude and is therefore beyond the reach of perturbative approaches. To provide a glimpse about the nature of the transition we considered the lowest-order Magnus expansion suggesting that the transition could correspond to the point where the kinetic term in the Floquet Hamiltonian becomes negative. 

The existence of a transition at a finite value of the driving amplitude is further supported by two numerical methods: a self-consistent time-dependent variational approach, and a semiclassical approach.
Interestingly, the emergent phase diagram (Fig.s \ref{fig:mehrtash} and \ref{fig:figure5}) is in qualitative agreement with a simple-minded quadratic expansion (Fig.~\ref{fig:phasediagram}). 
The dynamical phase diagram displays two distinct stability islands, respectively for large driving frequencies $\gamma$ and for large $g_0=1/T\int_0^T g(t)$. The former island becomes unstable when $\gamma$ is comparable to the short wavelength cutoff $\Lambda$. This suggests that the transition is of mean-field nature, thus bearing the same character as the parametric resonance of a single harmonic oscillator. The latter island is related to the existence of finite excitation gap, which protects the system from low-frequency drives. In both cases, we observe that quantum fluctuations promote ergodicity and decrease the value of the critical driving amplitude.

The observed transitions are analogous to the stability threshold predicted in Ref.s~ [\onlinecite{abanin_2014A,lazarides_2014B,ponte_bis_2014,abanin_2014B}] for many-body localized (MBL) states, but does not require localization in real space. Our findings are in contrast to the conclusions of D'Alessio \etal \cite{dalessio_2014}, who argued that generic many-body system should be dynamically unstable under a periodic drive.
Ponte \etal \cite{abanin_2014A} showed that ergodic systems are always unstable to the periodic modulation of a local perturbation: we find here that global coherent perturbations can display a qualitatively different behavior (See also Russomanno \etal  \cite{russomanno_2014} for a specific example).

\acknowledgments We are grateful to T. Giamarchi, B. Halperin, D. Pekker, A. Russomanno, K. Sengupta, A. Tokuno for many useful discussion.
The authors acknowledge the organizers of the KITP  workshop on ``Quantum Dynamics in Far from Equilibrium Thermally Isolated Systems'', during which
this work was initiated, and the NSF grant No. PHY11-25915. EGDT and ED thank the support of the Harvard-MIT CUA.
RC acknowledges the International Program of University of Salerno and the Harvard-MIT CUA. This research was supported by THE ISRAEL SCIENCE FOUNDATION (grant No. 1542/14).
AP and LD acknowledge the support of NSF DMR-1206410 and AFOSR FA9550-13-1-0039.

\appendix

\section{Third-order Magnus expansion}

\label{sec:magnus}

If the driving frequency $\g$ is the largest scale in the problem, Magnus expansion applies and going up to third order one has:
\begin{eqnarray}
 H^{(1)}_{\rm eff}&=&\frac{1}{T} \int_0^T dt \hat{H}(t_1)=\nonumber \\
&&=\int dx \lbrack K P^2 +\frac{1}{K} (\nabla \phi)^2-g_0 \cos(\phi) \rbrack \nonumber \\
 H^{(2)}_{\rm eff}&=&\frac{1}{2 T i} \int_0^T dt_1 \int_0^{t_1} dt_2 \lbrace\hat{H}(t_1),\hat{H}(t_2)\rbrace=0 \nonumber \\
H^{(3)}_{\rm eff}&=&\frac{1}{6 T i^2} \int_0^T dt_1 \int_0^{t_1} dt_2 \int_0^{t_2} dt_3 \nonumber \\
\times\left( \lbrace\hat{H}(t_1)\right.&,&\left.\lbrace\hat{H}(t_2),\hat{H}(t_3)\rbrace\rbrace+\lbrace\hat{H}(t_3),\lbrace[\hat{H}(t_2),\hat{H}(t_1)\rbrace\rbrace \right)\nonumber
\end{eqnarray}
where the time-integral domain is ordered $0<t_n<\ldots<t_2<t_1<T$ and $T$ is the period of the driving $T=\frac{2\pi}{\gamma}$.

In the present case the second-order term vanishes $H^{(2)}_{\rm eff}=0$: for time-reversal invariant perturbations, all even-order terms are exactly zero.
The third order Magnus expansion leads to the effective Hamiltonian (\ref{eq:Heff}), where we used the following identities:
\begin{align}
\lbrace P^2,\lbrace P^2,\cos(\phi)\rbrace\rbrace&= -2i \lbrace P^2, P\sin(\phi)\rbrace= 4 P^2\cos(\phi)\;;
\end{align}
\begin{align}
\lbrace(\partial_x\phi)^2,\lbrace P^2,\cos(\phi)\rbrace\rbrace&= -2i \lbrace(\partial_x\phi)^2, P\sin(\phi)\rbrace\nonumber \\&=2 (\partial_x\phi)\left[\partial_x\sin(\phi)+\sin(\phi)\partial_x\right]\;;
\end{align}

\begin{align}
\frac1{12\pi}\int_0^{2\pi} dt_1 \int_0^{2\pi} dt_2 \int_0^{2\pi} dt_3&\nonumber\\ \left[g(t_3)-g(t_2)\right.&\left.+g(t_1)-g(t_2)\right]  = g_1\;;\\
\frac1{12\pi}\int_0^{2\pi} dt_1 \int_0^{2\pi} dt_2 \int_0^{2\pi} dt_3 &\nonumber\\ g(t_1)[g(t_3)-g(t_2)]+g(t_3)&[g(t_1)-g(t_2)] = g_0 g_1 - \frac14g_1^2\;.
\end{align}

\section{Single Pendulum Limit}
\label{app:single}
\begin{figure}[h]
\center
\includegraphics[scale=0.45]{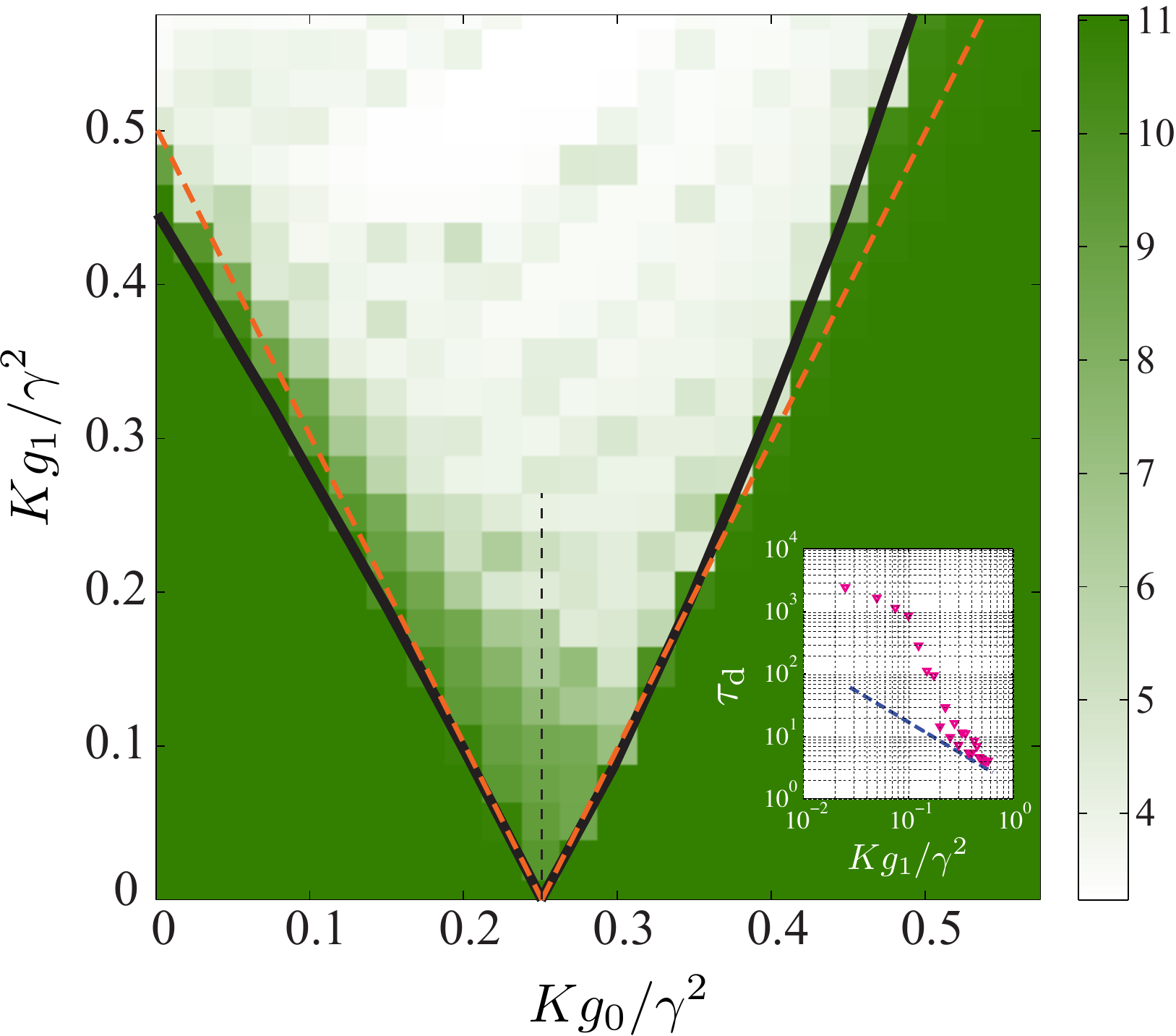}
\caption{Stability diagram obtained from the self-consistent variational approach for $\Lambda/\gamma=0.01$, $K=0.05\pi$. The color code shows $log(\tau_d)$, where $\tau_d$ is defined as $Z(\tau_d)/Z(0) = 10^{-3}$ (dark green areas correspond to $\tau_d>T_f=10^4 (2\pi/\gamma)$). Orange dashed lines are from the quadratic expansion and solid black lines are from Broer et al \cite{broer}. The inset plot shows $\tau_d$ as a function of $K g_1/\gamma^2$ along the resonance $Kg_0/\gamma^2 = 1/4$. The dashed line refers to $\tau_d/\sim(Kg_1/\gamma^2)^{-1}$.}
\label{fig:mehrtash2}
\end{figure}

As mentioned in the text, the limit $\Lambda\to0$ of Eq.(\ref{eq:H_in}) recovers the case of an isolated classical Kapitza pendulum. In Fig.~\ref{fig:mehrtash2} we show that this limit is correctly reproduced by the self-consistent variational approach. With decreasing $g_1$, the time required for the system to become unstable grows approximately as $1/g_1$. This explains the apparent inconsistency between the quadratic expansion and the self-consistent variational approach (Fig.~\ref{fig:mehrtash}(a)) for $\Lambda/\gamma\ll1$ and $K g_1/\gamma^2$.

\section{Finite time scaling}
\label{app:scaling}
\begin{figure}[h]
\includegraphics[scale=0.5]{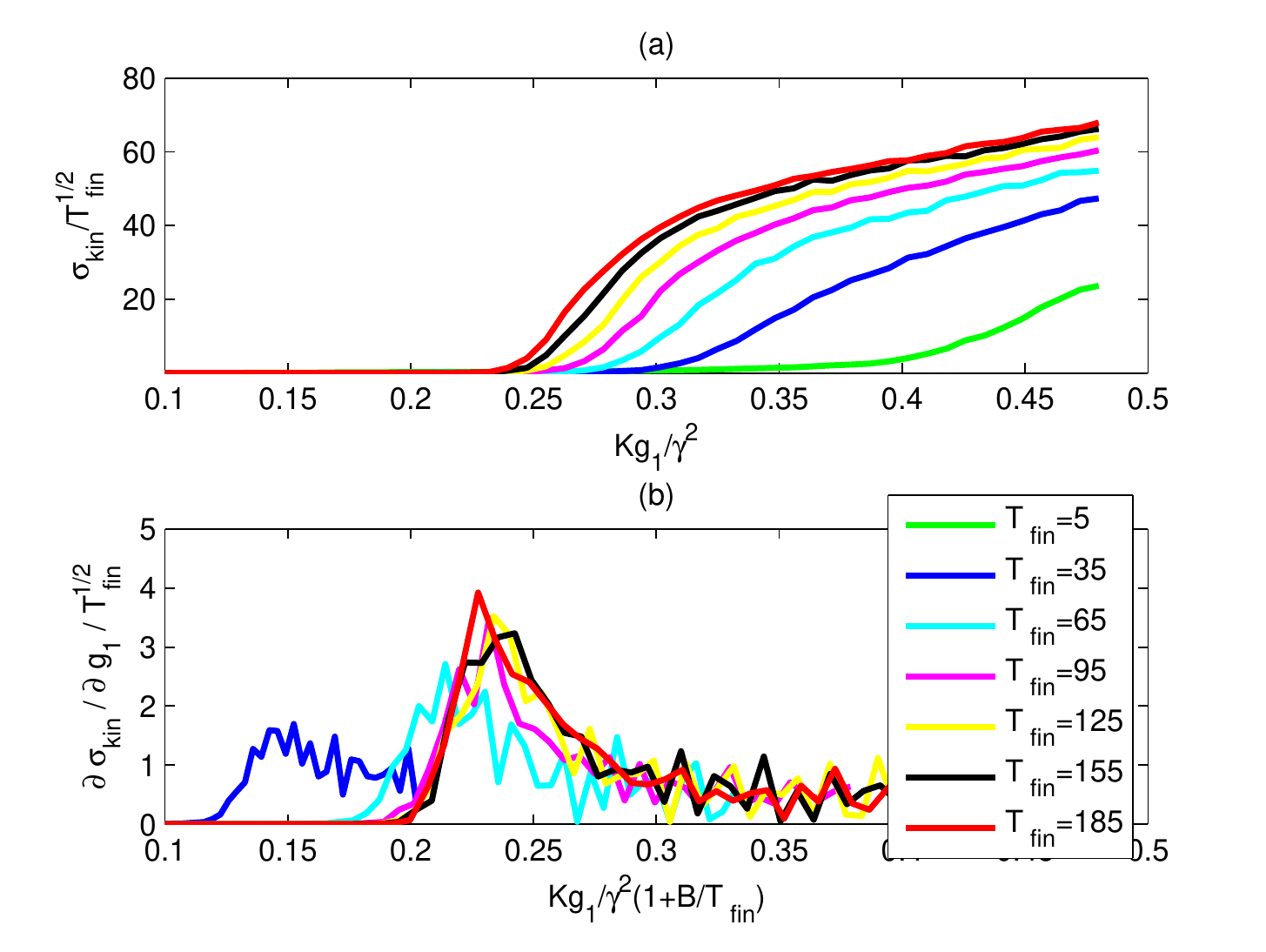}
\caption{Energy absorption rate as a function of the normalized periodic amplitude, for a different number of periods ($T_{fin}$). (a) raw data for a system of size $L=400$, $K=0.4\pi$ and $\Lambda/\gamma=0.1$; (b) same data on a normalized x-axis, showing a good data collapse when the position of the peak is rescaled as $g = g_c + A/T$ with $A=20$.}\label{fig:collapseA}
\end{figure}

\begin{figure}[t!]
\includegraphics[scale=0.5]{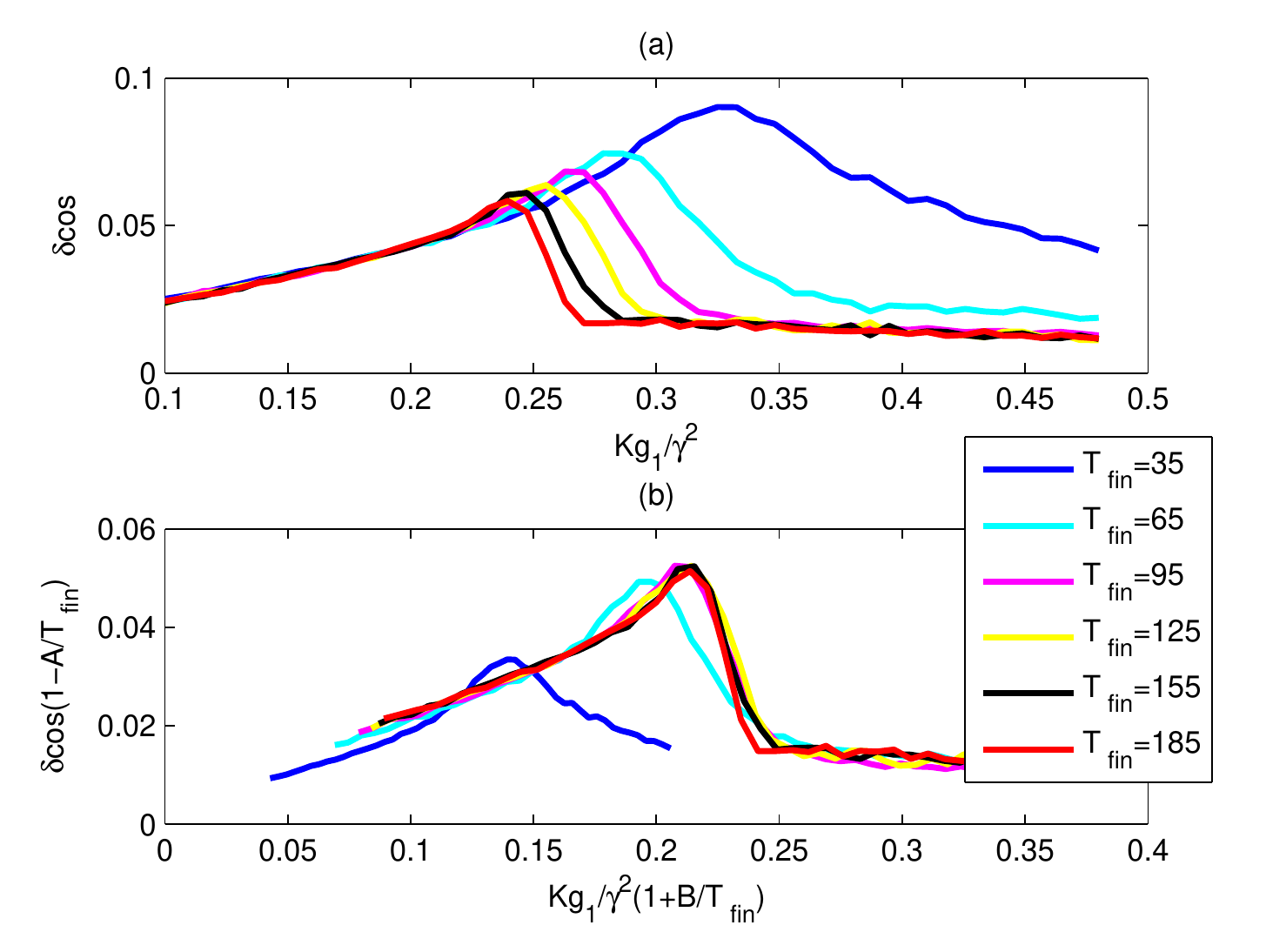}
\caption{Same as Fig.~\ref{fig:collapseA} for the amplitude of the cosine oscillations.}\label{fig:collapseB}
\end{figure}

In this Appendix we extend the analysis of Sec.~\ref{sec:TWA} and determine the scaling of physical observables as a function of the number of driving periods. Figs.~\ref{fig:collapseA}(a) and \ref{fig:collapseB}(a) show the energy absorption rate and the average cosine as a function of the driving amplitude, at different times. These plots show that all the curves tend to a well defined long-time limit. To better highlight this asymptotic limit, we shift each curve to take into account the dependence of the critical amplitude on $T_{\textrm fin}$  (Fig.~\ref{fig:asymptotic}). Specifically we consider finite-time corrections of the type: $g_c \to g_c (1+A/T_{\textrm fin})$. Through this transformation we obtain an excellent data collapse on a single universal asymptotic curve, as shown Fig.s \ref{fig:collapseA}(b) and \ref{fig:collapseB}(b). In both cases, the asymptotic curves become steeper and steeper as the number of driving periods increases. This may indicate that physical observables are ultimately not continuous in the $T_{\textrm fin}\to \infty$ limit, with important consequences for the universal properties of the transition.

\newpage


\begin{thebibliography}{90}
\bibitem{note0} In the thermodynamic limit, the many-body energy is always unbound.
\bibitem{note} Let us note that the Magnus expansion is valid in the limit of $g_0/\gamma^2 \ll 1$, leading to the additional requirement that
$g_1\gg g_0$, thus one can approximate $\tilde g$ with $\tilde g=\frac{K g_1^2}{4 \gamma^2}$.
\bibitem{note1} The driving amplitude must be small in any case to prevent chaotic behavior.
\end{thebibliography}
\end{document}